\documentclass{sig-alternate}
\usepackage{times,amsmath,epsfig}

\usepackage{cite}
\usepackage{amssymb}
\usepackage{graphicx}
\usepackage{wrapfig}
\usepackage{url}
\usepackage{listings}
\usepackage{amsmath} 
\usepackage{color}
\usepackage{algorithm}
\usepackage[noend]{algpseudocode}

\newfloat{program}{thp}{lop} 
\floatname{program}{Program}
\newcommand{\ie}{i.e.} 
\newcommand{\eg}{e.g.} 
\newcommand{\et}{et al. }
\newtheorem{mydef}{Definition}
\newtheorem{mytheorem}{Theorem}

\begin{document}

\conferenceinfo{}{}
\CopyrightYear{2014} 
\crdata{978-1-4503-2627-8/14/07\$15.00.\\
http://dx.doi.org/10.1145/xxxx.xxxx} 

\title{Optimizing ETL Dataflow Using Shared Caching and Parallelization Methods}

\numberofauthors{1}
\author{
\alignauthor
Xiufeng Liu\\
       \affaddr{University of Waterloo, Canada}\\
       \email{xiufeng.liu@uwaterloo.ca}
}

\maketitle
\begin{abstract} 
Extract-Transform-Load (ETL) handles large amount of data and manages workload through dataflows. ETL dataflows are widely regarded as complex and expensive operations in terms of time and system resources. In order to minimize the time and the resources required by ETL dataflows, this paper presents a framework to optimize dataflows using shared cache and parallelization techniques. The framework classifies the components in an ETL dataflow into different categories based on their data operation properties. The framework then partitions the dataflow based on the classification at different granularities. Furthermore, the framework applies optimization techniques such as cache re-using, pipelining and multi-threading to the already-partitioned dataflows. The proposed techniques reduce system memory footprint and the frequency of copying data between different components, and also take full advantage of the computing power of multi-core processors. The  experimental results show that the proposed optimization framework is 4.7 times faster than the ordinary ETL dataflows (without using the proposed optimization techniques), and  outperforms the similar tool (Kettle).
\end{abstract}

\section{Introduction}
\begin{figure}
  \begin{center}
    \includegraphics[width=0.35\textwidth]{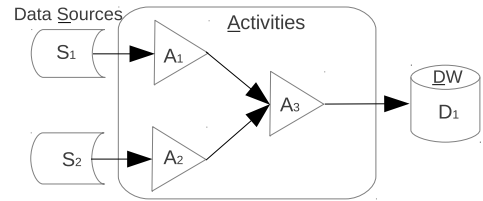}
  \end{center}
  \vspace{-20pt}
 \caption{Dataflow components of ETL}
\label{fig:etldataflow}
\vspace{-10pt}
\end{figure}

In data warehousing, ETL technology  is a collection of tools responsible for extracting, cleansing, customization, reformatting, integration and loading data from different sources into a data warehouse. Dataflow is a term used in computing architectures where a number of computing units are organized  logically in order to do the required computations in terms of user-defined  rules \cite{Sharp1992}. The concept of dataflow is introduced in the ETL process to define the data movement and transformation logic from sources to a central data warehouse. Figure~\ref{fig:etldataflow} shows a typical ETL dataflow in a data warehouse. It consists of the following four components: 1) data sources, \eg, operational databases, text files and so on; 2) data processing activities;  3) the data warehouse; and 4) the dependency of the activities which is denoted by the dashed arrows \footnote{The ETL activity occurs in a dataflow component. In the rest of this paper, we refer {\em component} and {\em activity} as an exchangeable concept.}.

Today, with the ever-growing volume of data, data warehousing  is facing the increasing pressure to enhance their data processing capabilities in that the ETL scenarios in many enterprises may take several hours or even days to complete. The delay in data processing might lead to in-accurate business decision-making.  Business intelligence (BI) developers therefore tend to improve their data warehousing system, e.g., to optimize their ETL dataflow which is one of the most commonly used techniques under existing IT infrastructures. The dataflow optimization techniques include dataflow designing, caching, parallelizing, job scheduling and so on. With an optimized ETL dataflow, the changes to data in source systems can be synchronized with the data warehouse in a timely manner. 
In the past few years, ETL dataflow optimization has received a growing attention from the research community, such as \cite{han2004,simitsis2010,simitsis2005,simitsis20051,Dayal2009,Tziovara20051}. Nevertheless, optimizing an ETL dataflow is a non-trivial task as it involves multiple aforementioned aspects. For example, in order to find an optimal dataflow design, the search space could grow exponentially \cite{simitsis2010}. Most existing ETL tools only provide the basic functionality for the optimization (such as re-ordering of the activities in ETL dataflow), however they do not provide more advanced features. 

\vspace{-1pt}
In an ETL dataflow, data is transfered and processed through a number of connected ETL components. In order to explain the ETL dataflow, we consider a simple scenario consisting of  two connected components (see A and B in Figure~\ref{fig:separatecache}). The component A processes the data and saves the results into an output cache. The results from the output cache are then transferred to an input cache of the component B by data copying operation. Since copying involves physical data movement and memory synchronization, it is  relatively expensive  in the dataflow. Furthermore, the activity thread of the component B might be inactive during the copying process, which may delay the data processing overall. In this paper, we propose the optimization framework that uses a shared cache  to remedy the data copying operation. In term of the characteristics of the activities, the optimization framework partitions a dataflow into several subsets, each of which contains one or more components.  The framework optimizes a subset by re-using a single cache to transfer the data, and processes the components in the cache in a sequential order. To reduce the processing time of the activities, the optimization framework makes use of the parallelization techniques including pipelining and multi-threading based on different granularities (subset-level or component-level within a subset). 
This proposed technique greatly maximizes the usage of the computing resources for a dataflow, e.g., running on a powerful machine. The proposed framework applies the optimization techniques at the transformation layer of ETL, whilst it remains transparent to BI developers.

In this paper, we make the following contributions. First, we classify the components in an ETL dataflow into three different categories based on their data processing characteristics: row-synchro\-nized component, semi-block component and block component. Second, we propose the partitioning algorithm to divide the dataflow based on these categories. The partitioning granularities (subset-level or component-level within a subset) provide the foundation to determine an appropriate parallelization method. Third, the proposed framework provides the built-in support for selecting appropriate optimization techniques including pipelining and multi-threading at different granularities. Fourth, we propose the concept of shared caching scheme (used within multiple ETL components) to reduce memory footprint and the algorithm to estimate the optimal degree of parallelization. Finally, we show the empirical evidence to verify the effectiveness of using the proposed optimization framework.

The paper is organized as follows. Section~\ref{sec:problem} describes the problems and the optimization framework. Section~\ref{sec:optanddataflow} proposes the optimization caching scheme and categorizes the ETL components in a dataflow. Section~\ref{sec:parallelstrategies} presents the dataflow partitioning and the parallelization methods used for the partitioned dataflows. Section~\ref{sec:evaluation} presents the implementation and the evaluation. Section~\ref{sec:relatedwork} discusses related work. Section~\ref{sec:conclusionandfuturework} concludes the paper and presents the future work.

\section{Problems and the ETL Optimization Framework}
\label{sec:problem}
\vspace{-5pt}
\begin{mydef}[ETL dataflow]
An ETL dataflow is formalized as a directed acyclic graph (DAG) \cite{simitsis2005}, $G(V, E)$,  where $V$ is a list of activities $A$ over the row set $R$, \ie, $V=A \cup R$, and $E$ represents as the set of logical transitions from an activity to the other activities.
\end{mydef}

\begin{figure}
\begin{center}
\includegraphics[width=0.42\textwidth]{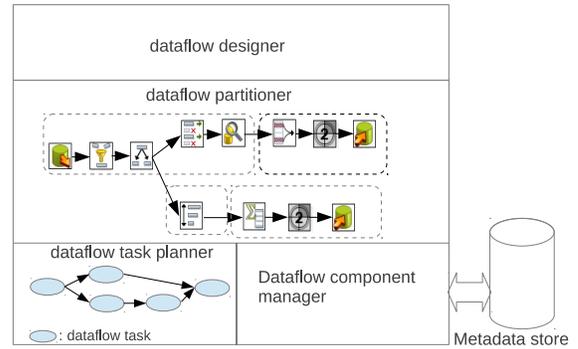}
\caption{ETL dataflow optimization framework}
\end{center}
\label{fig:optframework}
\end{figure}


In the dataflow, the row sets $R$ go through the components, and are processed by the activity on each component, \ie, does data transformation (This resembles to the work flow in the assemble line of a factory). In the dataflow, a cache is used to hold the data temporarily between two continuous neighbor components. In the entire dataflow, the caches could cause a lot of memory consumption for holding the intermediate data. Moreover, since the data in a cache is passed to a downstream component as the input, it triggers the copying operation, which requires significant CPU resources. The proposed optimization framework remedies this by using shared caching scheme, where a single cache is re-used by the neighbor components. Therefore, no cache copying is needed and the consumption of CPU resource for copying is also saved. In shared caching scheme, an ETL dataflow is first partitioned based on the types of components (we will discuss the partitioning in Section~\ref{sec:parallelstrategies}),  then shared caches are used in each of the partitions.

The system architecture of the optimization framework is presented in Figure~\ref{fig:optframework}. It consists of the following five functional modules:
\begin{itemize}
	\item {\em Dataflow designer}: This module is used to design an ETL dataflow with the toolkits offered by the system. The toolkits include the utilities for designing work such as update, edit, search and replace, etc., and the components for building a dataflow including different type of data sources, targets, and the components for different ETL transformation operators. The dataflow designer provides a graphic user interface in which programmers can conveniently implement a dataflow according to user requirements.
	\item {\em Dataflow partitioner}: This module  is responsible for partitioning a dataflow based on the types of dataflow components. The optimization techniques are applied in each partition to improve the execution concurrency of the dataflow tasks. 
	\item {\em Dataflow task planner}: This module  is responsible for planning dataflow tasks automatically. When a dataflow is partitioned, the job for dataflow is thus generated into multiple tasks, and task planner will plan the execution order according to the dependency of the generated tasks.
	\item {\em Dataflow component manager}: This module  is responsible for  data processing and data source component management. It is in charge of component life cycle management and component specification management.
	\item {\em Metadata store}: The store  is responsible for the metadata management. The metadata includes the schema information of data sources and data processing components, the specifications of  ETL dataflows and the information of job and task planning. Metadata can be imported from or exported to XML files.
\end{itemize}

\section{Optimize Caching Scheme}
\label{sec:optanddataflow}
\begin{figure*}[htp]
\begin{minipage}[b]{0.33\linewidth}
\centering
\epsfig{file=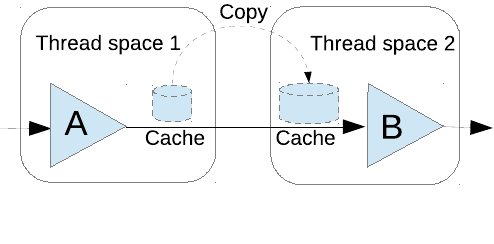, scale=0.85}
\vspace{-10pt}
\caption{Separate cache}
\label{fig:separatecache}
\end{minipage}
\begin{minipage}[b]{0.33\linewidth}
\centering
\epsfig{file=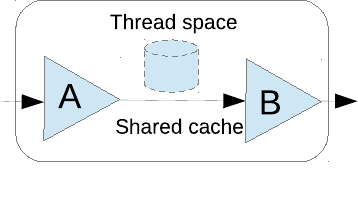, scale=0.85}
\vspace{-10pt}
\caption{Shared cache}
\label{fig:sharedcache}
\end{minipage}
\begin{minipage}[b]{0.3\linewidth}
\centering
\epsfig{file=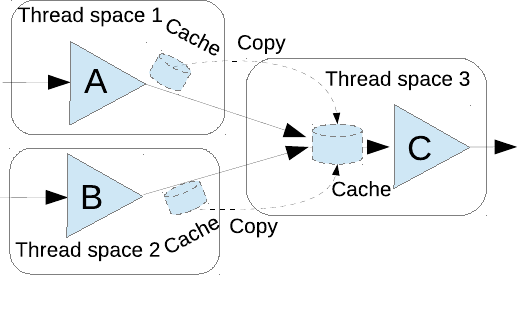, scale=0.79}
\vspace{-15pt}
\caption{Separate cache}
\label{fig:semiblockcomp}
\end{minipage}
\end{figure*}

This section illustrates the traditional approach of transferring data between two components shown as  A and B in Figure~\ref{fig:separatecache}.  If we use separate caches for the output and input, the extra memory and CPU resources are needed to transfer the data (shown as \texttt{Copy}). However, in some cases the cache can be re-used for optimization, even by a number of connected components. We call this as {\em shared caching scheme} (see Figure~\ref{fig:sharedcache} which only shows two components for simplicity). 

\begin{figure*}[htp]
\centering
\includegraphics[scale=1]{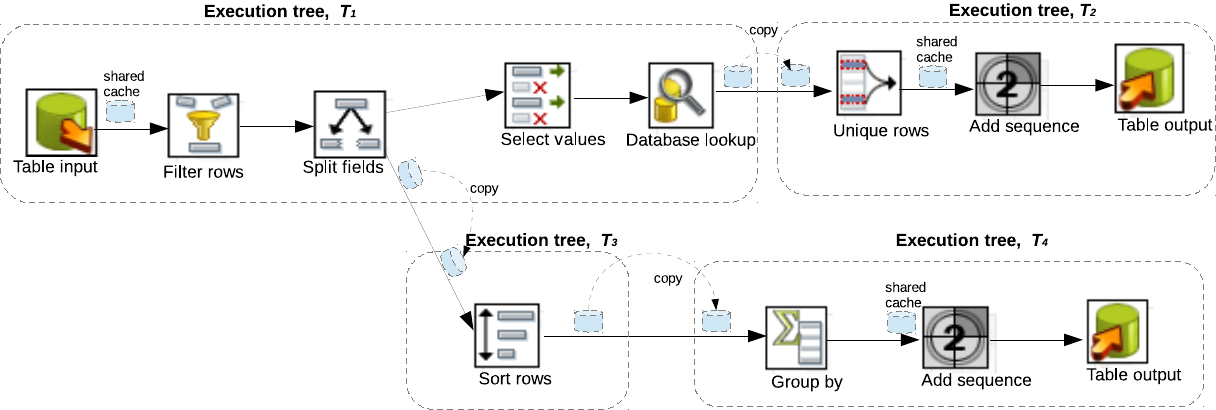}
\caption{Dataflow partitioning}
\label{fig:paritionexetree}
\end{figure*}

Before we apply the shared caching scheme, we first classify the components into the following three categories (the types of the components are saved in the metadata store):

\begin{itemize}
 \item {\bf Row-synchronized Components:} This component performs row-based data processing, such as \texttt{filter}, \texttt{lookup},\linebreak \texttt{splitter}, \texttt{data format converter}, etc. It processes rows one after the other. In this component, the shared caching scheme is applied to optimize the component activities using a single thread space (see Figure~\ref{fig:sharedcache}).
 
\item {\bf Block Components:} This component receives the rows from a single upstream component. Data processing cannot be started until all the rows have been received, \eg, the component B in Figure~\ref{fig:separatecache}. The output of component A is added into the input cache by copying operation. The components in this category are mainly consists of aggregation operators, such as \texttt{avg}, \text{sum}, \texttt{min} and \texttt{max}, etc. Each activity of a block component runs its own thread. Block component has to accumulate all the rows that is why the block components are the least efficient. 
 
 \item {\bf Semi-block Components:} This component receives the rows from multiple upstream components. Some what similar to the previous component, data processing cannot be started until all the rows that satisfy a certain condition have been received, \eg, the component C in Figure~\ref{fig:semiblockcomp}. The rows from its upstream components are added into the input cache by copying operation. The semi-block components consist of \texttt{union}, \texttt{merge} and \texttt{alike}.
\end{itemize}

\section{Dataflow Partitioning and Parallelization}
\label{sec:parallelstrategies}
In order to optimize, we first perform the vertical partitioning by dividing a dataflow into multiple subsets, or {\em sub-dataflows},  then we use parallelization technology in each of the subsets. When we do the partitioning, the partitioning granularity  should not be  either too coarse or too fine. If the granularity is too coarse, the execution concurrency of the sub-dataflows will become very low, which is not able to take full advantage of the computing resources, if running on a powerful machine. On the contrary, if the granularity is too fine, it will lead to a lot of concurrent threads, and cause overall performance loss for competing of the computing resources. In addition, data synchronization between two neighbor sub-dataflows could also lead to some overhead, and so does thread switching when multi-threading is used. Based on these considerations, we therefore propose the {\em 3-level} partitioning method:  first vertically partition a dataflow into multiple sub-dataflows, called {\em execution trees} (the {\em coarse-level}), then  horizontally partition the input of an execution tree ({\em medium-level}), and finally do a further partitioning in the staggering components of an execution tree ({\em fine-level}). For each partition with a different granularity, we choose an appropriate parallelization method to optimize it (if applicable).

\subsection{Execution Tree}
\begin{mydef}[Execution Tree] An execution tree is a directed acyclic graph, defined as $T(V', E')$. It is a subgraph of the  dataflow graph $G(V, E)$, where $V'$ is a nonempty subset of $V$ and $E'$ is a
subset of $E$. In $T$, the vertex with indegree equal to zero is the $root$, and the vertex with zero outdegree is the leaf. 
\end{mydef}

Suppose that we have done the coarse-level partitioning for the ETL dataflow graph $G$, which results in $n$ execution trees, $T_1, T_2,$ $ ..., T_n$. The generated execution trees also form a graph, defined as $G_{\tau}(V_{\tau}, E_{\tau})$, where $V_{\tau} = \{T_1, T_2,$ $ ..., T_n\}$, and $E_{\tau}$ is the set of the edge of any two connected execution trees. $G_{\tau}$ is a directed acyclic graph.

We now illustrate the coarse-level partitioning using the example shown in Figure~\ref{fig:paritionexetree}. The example depicts the ETL dataflow of loading data from a single data source into two target tables through a number of ETL transformation operators. As discussed earlier (Section~\ref{sec:optanddataflow}) that block and semi-block components require their own caches to accumulate the data before processing,  this dataflow can be partitioned into four execution trees, $T_1, T_2, T_3$ and $T_4$, each of which starts from the root, such as a data source, semi-block or block component. In an execution tree, the output of the root is horizontally partitioned into multiple splits, each of which is hold in a shared cache, and  passed to the downstream row-synchronized components. For any two connected execution trees, a new cache is needed, and the data is transfered to the new cache by COPY, \eg,  $T_1 \rightarrow T_2$, $T_1 \rightarrow  T_3$, and $T_3 \rightarrow  T_4$.

Algorithm \ref{alg:workflowpartitioning} shows how to partition a dataflow. It takes a dataflow graph,  $G$, as the input, and returns an execution tree graph, $G_{\tau}$, as the output. The algorithm does the partitioning using depth-first searching (DFS), which starts from the data source components, \ie, the vertexes whose in-degree is equal to zero in $G$ (see line 6--9).  An execution tree is created taking a data source component as the {\em root} (see line 7).  The algorithm then does the depth-first searching the downstream components, and adds them as the children of the tree. If a component is neither block type nor semi-block type, it will be regarded as a child added to the tree (see line 15); Otherwise, the searching for downstream components will finish.  A new execution tree is created, which takes this block or semi-block component as the root  (see line 17), and the  tree is added into the execution tree graph, $G_{\tau}$ (see line 18--19). At the end of the algorithm,  the graph, $G_{\tau}$, is returned,  containing one or several partitioned execution trees (the vertexes). For each of the execution trees, we optimize it using the shared caching scheme. All the threads for the activities use a single cache, thus data copying is not needed any more.
\begin{algorithm}[htp]
\caption{Partition an ETL dataflow graph, $G$}
\algdef{lS}{emit}{\textkeyword{EMIT}}%
{\scriptsize
\begin{algorithmic}[1]

\Function {partition}{$G$}
	\State	$G_{\tau} \gets$ \textproc{InitEmptyGraph}$()$
	\ForAll {$v \in V(G)$}
		\State $visited[v] \leftarrow 0$
	\EndFor
	\ForAll {$v \in V(G)$}
		\If {\textproc{InDegree}$(v) = 0 $ {\bf and} $visited[v] = 0$}
			\State $T \gets$ \textproc{CreateExecutionTree}$(v)$ \Comment{$v$ is the $root$ in $T$}
			\State $V(G_{\tau}).$\textproc{add}$(T)$ \Comment{Add $T$ as a vertex in graph $G_{\tau}$}
			\State \textproc{dfs}($G,  G_{\tau}, v, T$)
		\EndIf
	\EndFor
 \Return $G_{\tau}$	
\EndFunction

\Function {dfs}{$G,  G_{\tau}, v, T$}
	\State $visited[v] \gets 1$
	\ForAll {$u$ adjacent to $v$ }
		\If{$visited[u]=0$}
			\If{\textproc{type}$(u)\neq$`block' {\bf and} \textproc{type}$(u)\neq$`semi-block'}
			   \State $v.children.$\textproc{add}$(u)$
			\Else
				\State $T' \gets$ \textproc{CreateExecutionTree}$(u)$
				\State $V(G_{\tau}).$\textproc{add}$(T')$ 
				\State $E(G_{\tau}).$\textproc{add}$(T\rightarrow T')$ \Comment{Add the edge $T\rightarrow T'$}

			\EndIf
			\State $visited[u] \gets 1$
			\State \textproc{dfs}$(G, G_{\tau}, u, T)$
		\EndIf
	\EndFor
\EndFunction
\end{algorithmic}
}
\label{alg:workflowpartitioning}
\end{algorithm}

\subsection{Inside Execution Tree Parallelization}
\begin{figure}[ht]
\centering
\includegraphics[scale=1.5]{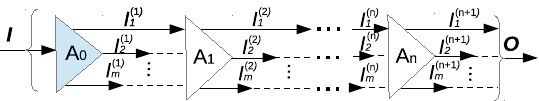}
\caption{Pipeline parallelization in an execution tree}
\label{fig:sharedcacheinexetree}
\end{figure}

\begin{mydef}[Inside Execution Tree Parallelization] 
The output of the root of an execution tree is horizontally partitioned, and pipeline parallelization is used within the execution tree to process the partitions.
\end{mydef}
\vspace{-5pt}
We now give the formalization. The ETL process of an execution tree is  a 3-tuple $(\mathcal{I}, \mathcal{F}, \mathcal{O})$ where $\mathcal{I}$ represents the input, $\mathcal{F}$ represents the ETL activities of the components, and $\mathcal{O}$ represents the output of the execution tree (see Figure~\ref{fig:sharedcacheinexetree}). 
We use  the tuple, $(A_0, A_1, ..., A_n)$,  to denote the activities of the components in an execution tree. The function $f_0:\mathcal{I}\rightarrow \Sigma$ is used to denote the activity of the first component, $A_0$, where $I$ represents the input of the execution tree, and $\Sigma$ represents the output of $A_0$. We horizontally partition  $\Sigma$ into $m$ even splits (the value of $m$ is configurable), \ie, $\Sigma=(I_1^{(1)}, I_2^{(1)},  ...,I_m^{(1)})$, where $I_i^{(1)}, i=1...m$ disjoints the subsets of $\Sigma$. Here, we use,  $I_i^{(j)}$, to denote the $i$th horizontal split as the input for the $j$th activity, and  use $|I_i^{(j)}|$ to denote the number of rows for the split. 
The final output of an execution tree is $\mathcal{O}=(I_1^{(n+1)}, I_2^{(n+1)},  ...,I_m^{(n+1)})$ where $|I_i^{(n+1)}|\geq 0, i=\overline{1, m}$. 

 An activity, $A_j$, can be expressed as a function, \ie, $f_j: I_i^{(j)}\rightarrow I_i^{(j+1)}$ where $i=\overline{1, m}$ and $j=\overline{1, n}$. Thus, the activities $(A_0, A_1,$ $ ..., A_n)$ can be formalized as $\mathcal{F}(x)=f_n(f_{n-1}(...f_1(x)))$, $x \in \Sigma$, which is made of the activity functions recursively. 


\begin{figure}
  \begin{center}
    \includegraphics[width=0.4\textwidth]{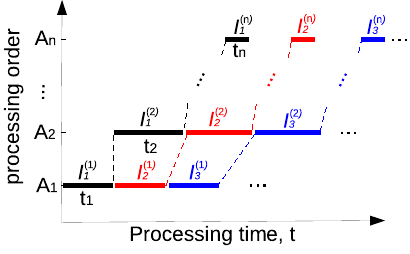}
  \end{center}
  \vspace{-10pt}
\caption{Pipeline parallelization}
\label{fig:assemblyprocessing}
\end{figure}
A shared cache is created to hold each of the splits, and the same cache will be pass through all the downstream activities, each of which runs a separate thread. Therefore, $n$ activity threads run in parallel in an execution tree   (see Figure~\ref{fig:sharedcacheinexetree}). In pipeline parallelization, when a component has finished processing a shared cache, it passes it to the next immediately. Algorithm~\ref{alg:pipelineparallelism} describes the pipeline parallelization with a degree $m'$. In order to limit the memory usage, we set the value of $m' \leq m$. We exploit a fix-sized blocking queue, and a house keeping thread to maintain the maximal number of parallel threads, and the memory usage (see line 14-15). A shared cache is created to hold each of the splits from the first component (of block or semi-block type, see line 17--18). The \emph{pipeline consumer thread} is created to process the shared cache (see line 18--19). The thread will be blocked if there is no space available in the queue. The shared cache is processed sequentially through all the activity threads  (see line 1--11).  For each component, it has to process $m$ shared caches in a sequential order. Therefore, if a component is processing a shared cache, and a pipeline consumer thread passes it a new shared cache, the pipeline consumer thread will be halted by \texttt{wait} (see line 7). It will not be waken up until the component has finished its processing (by \texttt{notifyAll}, see line 11).

When a shared cache is processed by the component, $A_i$, the total processing time, denoted by $t_i$, consists of the time for processing all rows, and miscellaneous time, denoted by $t_0$. The miscellaneous time is consumed by any other necessary actions, such as create and clean temporary tables, and hand  over a shared cache to other components, etc.,  typically, $(t_i-t_0) \gg t_0$. We call  $(t_i-t_0)$ as net time for processing the rows in activity $A_i$.  Therefore, when a shared cache goes through all the $n$ sequential activities, the total time is $\sum_{i=1}^{n}{t_i}$ (see Figure~\ref{fig:assemblyprocessing}). For simplicity, we now consider when the $m$ splits can fit into the memory, and use the same number of pipelines to process the splits, \ie, $m=m'$. Suppose that the activity $A_j$ has the maximal time, \ie, $t_j=max(t_1, t_2, ..., t_j, ..., t_n)$,  according to the pipeline workflow algorithm in \cite{han2004}, the total time is $T_{p}=\sum_{i=1}^{j-1}t_i + mt_j+\sum_{i=j+1}^{n}t_i$.

\begin{algorithm}[htp]
\caption{Pipeline parallelization within an execution tree}
\algdef{lS}{emit}{\textkeyword{EMIT}}%

{\scriptsize
\begin{algorithmic}[1]
\algblockdefx[NAME]{START}{END}
   [2][Unknown]{{\bf class} #1}
{}

\START[PipelineConsumerThread]

\Function {Init}{ActivityThreads $\Gamma$, SharedCache $sc$}
	\State $\Gamma', sc' \gets \Gamma, sc$
\EndFunction
\Function {run}{\ }
	\ForAll{$a \in \Gamma'$}
		\While{$a.busy$}
			\State $a.wait()$ \Comment{Waken up when $a$ has finished processing the previous shared cache}
		\EndWhile
		\State $a.busy \gets true$
	 	\State $sc' \gets a.process(sc')$ \Comment{Process a shared cache}
	 	\State $a.busy \gets false$
	 	\State $a.notifyAll()$ \Comment{Notify all the awaiting pipeline consumer threads if any}
	\EndFor
\EndFunction
\END

\Function{pipelineParallelization}{activity threads $\Gamma(a_1, ..., a_n)$, no. of splits $m$, parallel degree  $m'$}
		\State q $\gets$ new BlockingQueue($m'$) \Comment{Create a blocking queue with the specified capacity $m'$}
		\State new HouseKeepingThread($q$).start() \Comment{This thread is used to clean the invalid objects in $q$}
		\For {$i=0$ to $m-1$}
		\State $sc \gets $ Create a new shared cache
		\State Read a split of the output of the first component into $sc$
		\State $th \gets $ {\bf new} pipelineConsumerThread($\Gamma, sc$)
		\State $q.add(th)$ \Comment{The addition will be blocked if $q$ is full}
		\State $th.start()$
	\EndFor
\EndFunction
\end{algorithmic}
}
\label{alg:pipelineparallelism}
\end{algorithm}

If we process all the shared caches in a sequential order,  the total time, denoted by $T_s$, will become $m\sum_{i=1}^{n}t_i$. If we exclude the miscellaneous time of each activity, the total net time  can be regarded as a constant value, \ie, $c=m\sum_{i=1}^{n}(t_i-t_0)$ since the number of rows  processed in each activity is a constant. Therefore, $T_p$ can be represented as:
\vspace{-10pt}
 \begin{equation}
 T_p = \frac{c}{m} + (m-1)t_j + nt_0
 \end{equation}

\vspace{-10pt}
\begin{mytheorem}
For a fix-sized input of an execution tree, if the degree of pipeline parallelization is set to $m=\sqrt{\frac{c-\lambda N}{t_0}}$, the cost is minimal.
\end{mytheorem}
\vspace{-11pt}
\begin{proof}
{\em Suppose the total number of rows processed by the staggering activity, $A_j$, is $N$ (Note that $N$ is not necessary to equal to the size of the execution tree input $|\Sigma|$. For example, the upstream component of $A_j$ could be a {\em filter} operator for screening noisy data). The size of each horizontal split in $A_j$ is $\frac{N}{m}$, and we assume that the time used is linear to the split size. $t_j$ can be represented as $t_j = t_0 + \lambda\frac{N}{m}$ where $\lambda$ is an coefficient of the split size. Therefore,
\begin{flalign*}
 \begin{split}
& T_p = \frac{c}{m}+ (m-1)(t_0 + \lambda\frac{N}{m}) + nt_0\\
&  = \frac{c-\lambda N}{m} + t_{0}m + \lambda N + (n-1)t_0\\
& \geq 2\sqrt{t_{0}(c-\lambda N)} + \lambda N + (n-1)t_0
 \end{split}
\end{flalign*} when $ \frac{c-\lambda N}{m} = t_{0}m$, \ie, $m=\sqrt{\frac{c-\lambda N}{t_0}}$. The total time $T_p$ has the minimal value $2\sqrt{t_{0}(c-\lambda N)} + \lambda N + (n-1)t_0$. }
\end{proof}
\vspace{-10pt}

\begin{figure}[htp]
  \begin{center}
    \includegraphics[width=0.4\textwidth]{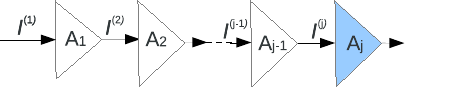}
  \end{center}
\caption{Identify the staggering activity, $A_j$}
\label{fig:mosttimconsumingactivity}
\end{figure}

From the discussion above, we can see the degree of parallelization is relevant to the size of the staggering component. But, it is obvious to know the range of the value, $1\leq m \leq |\Sigma|$. If  $m=|\Sigma|$, it means that we pipeline for every single row. It has the highest degree of parallelization. However, this case will lead to frequent transferring of the shared caches between activity threads, which dominate the overall time. On the contrary, if  $m=1$, the ETL workflow will degenerate to non-pipeline fashion (or sequential execution) since the input will be processed once for all through all the activities. To make an optimal optimization, we need to find staggering component, compute the degree of parallelization, then adjust all the activities. Algorithm~\ref{alg:findoptimal} describes how to find the staggering activity, \eg, the grayed one $A_j$ in Figure~\ref{fig:mosttimconsumingactivity}, and compute the optimal degree of parallelization.

\begin{algorithm}[htp]
\caption{Compute the optimal degree of parallelization with a given input, $\Sigma$}
\algdef{lS}{emit}{\textkeyword{EMIT}}

{\small
\begin{algorithmic}[1]
\Require The sample data set $D$ (with $m'$ horizontal splits), and the number of activities in an execution tree, $n$.
\State Measure the total miscellaneous time, $T_0$.
\State	Run an execution tree in non-pipeline fashion to process $m'$ splits, record the time of each activities, and measure the total processing time $T_s$.
\State Find the staggering activity, $A_j$, compute the constant $c \gets (T_s-T_0)$, and the average miscellaneous time, $t_0 \gets \frac{T_0}{n}$.

\State Run an execution tree in pipeline parallelization, and  process the $m'$ splits, and compute the coefficient $\lambda$.

\State According the formula in {\em theorem 1}, compute the degree of parallelization, $m$.
\end{algorithmic}
}
\label{alg:findoptimal}
\vspace{-2pt}
\end{algorithm}

In this algorithm, we first measure the total time without giving any input to an execution tree. This is to approximate the miscellaneous time when  processing data,  such as the time for initializing and thread switching, etc. (see line~1).  We, then, use the sample data (random selection on rows from the data to be processed) as the input to execute the dataflow in non-pipeline and pipeline fashion, respectively, and finally compute the value of $m$ according to the formula (line~2--5).

\subsection{Inside Component Parallelization}
\begin{mydef}[Inside Component Parallelization] 
 It refers to using multi-threading technology to parallelize data processing in a heavy-computation component of the dataflow.
\end{mydef}

\begin{figure}
\centering
    \includegraphics[width=0.40\textwidth]{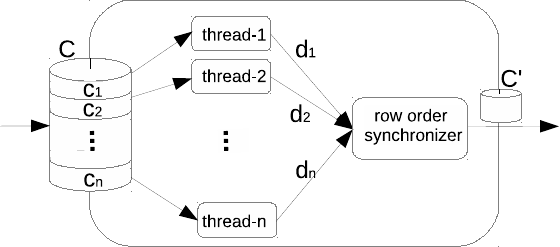}
\caption{Inside-component parallelization}
\label{fig:insidecompparallel}
\end{figure}

In an execution tree, the computation task of some components is much heavier than others, and forms the bottleneck. We make the further optimization of using multi-threading to parallelize data processing in a component. Multi-threading is applied to the non-block type components, \eg, the \texttt{Filter rows} component in $T_1$ of Figure~\ref{fig:paritionexetree}. Figure~\ref{fig:insidecompparallel} shows the scenario.  For a shared cache, $C$, delivered to it, the system firstly divides the rows in $C$ evenly into multiple splits, \ie, $C=\{c_1, c_2, ..., c_n\}$. Then, the component spawns a number of threads to process the $n$ splits in parallel (the value of $n$  can be set in the system configuration file), each of which results in an output $d_i, i=\overline{1...n}$. Finally, the outputs  $d_1, d_2, ..., d_n$ are merged by the \texttt{row order synchronizer}, which  maintains the row order of the output to be the same of the input (In some cases such as the activities \texttt{sort-filter-merge},  the input and output row order of \texttt{filter} should not be changed since the downstream activity is \texttt{merge}). The merged rows are saved into another shared cache $C'$, and delivered to the downstream component.

\section{Implementation and Evaluation}
\label{sec:evaluation}

We implement the optimization framework based on the open source ETL tool, Talend \cite{talend}, also known as Talend Open Studio for Data Integration with an intuitive, graphical, drag and drop design environment.  We add the extension, {\em dataflow partitioner},  to the transformation engine, and make it support the shared caching and pipeline parallelization.  The degree of pipeline parallelization of  execution trees can be configured in the configuration file. For the non-block type components including the \texttt{row filter, splitter, exploder, expression},  \texttt{lookup}, etc., they are extended to support the inside-component parallelization using multi-threading.  The number of threads is also configurable. If the number is not set, the system uses one as the default value,  which means that the inside-component parallelization is disabled. In addition,  XML is used as the metadata repository (RDBMS could also be configured to use), and the repository manager is extended to support managing the partitioning information of a dataflow.

The host of the experiments is a HP ProLiant DL380 G5 Server with two Intel Xeon quad-core processors (2.4GHz), 8GB RAM and a SATA hard disk (350 GB, 3 GB/s, 16 MB Cache and 7200 RPM), running Ubuntu 12.04 LTS with 64bit Linux 2.6.32 kernel.   JVM 1.6.0.29, and the option  ``-Xms2500m -Xmx6000m'' are used to run the program. PostgreSQL 8.4 is used as the data warehouse DBMS with the settings ``shared\_buffers=512MB, temp\_buffers= 128MB, work\_mem=56MB, checkpoint\_seg\-ments=20" and default values for other configuration parameters. 

We use the star schema benchmark (SSB) based on TPC-H \cite{ssd2007} for the evaluation. The star schema consists of one fact table, \texttt{lineorder}, and four dimension tables including \texttt{customer}, \texttt{part}, \texttt{supplier} and \texttt{date} (see \cite{ssd2007} for the detailed schema information). In the experiments, we use the fix-sized data sets for the  dimension tables: \texttt{customer} (150,000 rows, 725M), \texttt{part} (24,000 rows, 323M), \texttt{supplier} (231,000 rows, 150M) and \texttt{date} (10,000 rows, 134k), but vary the size of fact data. In all tests, we first load the  data sets into the data warehouse, then execute the dataflow corresponding to a SSB query, and finally write the query results into a text file.  The evaluation divides into two parts: the evaluation of the proposed optimization techniques, and comparing with other ETL tools.


\subsection{Evaluate the Framework}
We first use the query, Q4.1 (see the script bellow) to evaluate the optimization framework itself.  

{\scriptsize
\begin{verbatim}
SELECT d_year, 
       c_nation, 
       SUM(lo_revenue - lo_supplycost) AS profit 
FROM date, customer, supplier, part, lineorder 
WHERE lo_custkey = c_custkey AND lo_suppkey = s_suppkey AND
      lo_partkey = p_partkey AND lo_orderdate = d_datekey AND
      c_region = 'AMERICA' AND s_region = 'AMERICA' AND
      (p_mfgr = 'MFGR#1' or p_mfgr = 'MFGR#2') 
GROUP BY d_year, c_nation 
ORDER BY d_year, c_nation 
\end{verbatim}
}

\begin{figure}[htp]
\centering
\epsfig{file=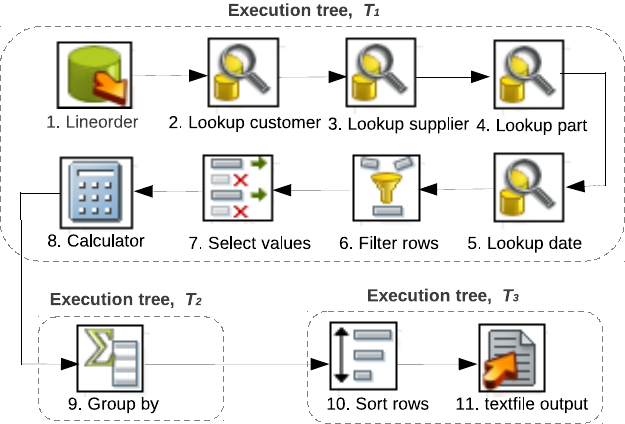, scale=1.2}
\caption{Partitioned dataflow for SSB Q4.1}
\label{fig:ssdexecutiontree}
\end{figure}

Figure~\ref{fig:ssdexecutiontree} shows the corresponding dataflow, which is partitioned into three execution trees using  Algorithm~\ref{alg:workflowpartitioning}. The dataflow reads the rows from the fact table \texttt{lineorder}, then  joined with the four dimension tables respectively by \texttt{lookup}. The join conditions are \texttt{lo\_custkey=c\_custkey AND c\_region ='AMERICA'}, \newline \texttt{lo\_suppkey=s\_suppkey AND s\_region = 'AMERICA'}, \texttt{lo\_partkey=p\_partkey AND (p\_mfgr='MFGR\#1' or \newline p\_mfgr='MFGR\#2')}, and \texttt{lo\_orderdate=d\_datekey}. \newline Component 6 filters the rows that fail to join with any of the dimensions (whose returned key values is equal to the default value, -1). Component 7 selects the necessary field values for the query by projection. Component 8 computes the value of the expression  \texttt{(lo\_revenue - lo\_supplycost)} of each row. Component 9 groups the computed values by \texttt{sum} operator, then component 10 sorts the values in each group, and component 11 writes the final results into a text file. The partitioned dataflow comprises three execution trees, $T_1, T_2$ and $T_3$.  $T_1$ consists of eight components, whilst  $T_2$ and $T_3$ contain one and two components, respectively.  $T_1$ has four \texttt{lookup}s, which are the relatively expensive operators in terms of the computing resource consumptions.

\begin{figure*}[htp]
\begin{minipage}[b]{0.5\linewidth}
\centering
\epsfig{file=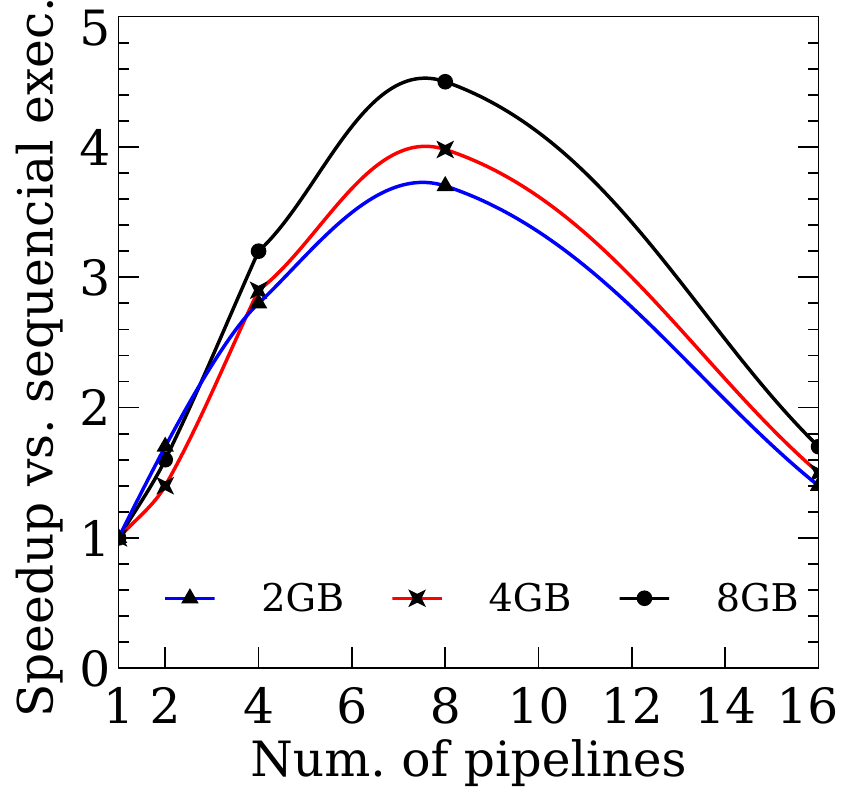, scale=0.5}
\caption{Speedup vs. num. of pipelines}
\label{fig:speedup_pipeline}
\end{minipage}
\begin{minipage}[b]{0.5\linewidth}
\centering
\includegraphics[scale=0.5]{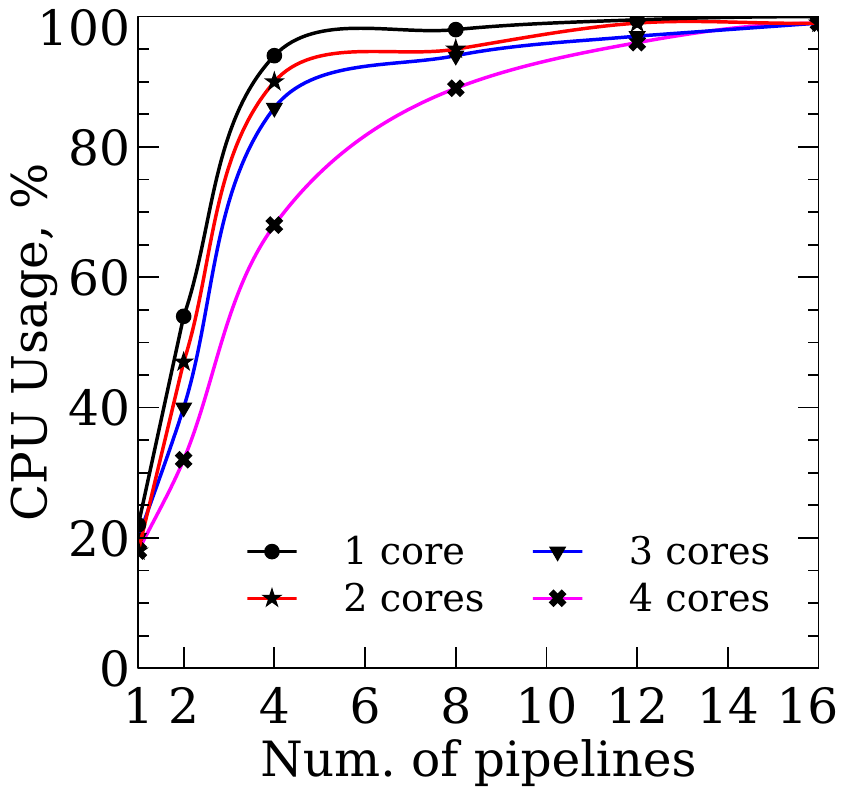}
\caption{CPU usage vs. no. of pipelines}
\label{fig:cpu_threads}
\end{minipage}
\end{figure*}

We first evaluate the speedup when the pipeline parallelization is applied to $T_1$. We execute the dataflow when the fact table is loaded with 2, 4, and 8 GB data sets, respectively.   Figure~\ref{fig:speedup_pipeline} shows the experimental results. The  speedup value is computed by dividing the sequential execution time (no pipeline) by the time of using pipelines. As shown, for each data set the speedup  scales nearly linear when the number of the pipelines is less than eight, but the speedup decreases dramatically when more pipelines are added. It is because more threads have been generated when the degree of pipelines is increased. This causes by  CPU bound (we show this in the next experiment). From  Figure~\ref{fig:speedup_pipeline}, we  also see that the speedup is more significant to  a bigger sized data set. The optimal number of pipelines is about eight, where the speedups are 4.7x, 3.9x and 3.7x  for the fact table loaded 2, 4, and 8 GB data, respectively. 

Figure~\ref{fig:cpu_threads} shows the CPU usage of executing the dataflow when the fact table is loaded 8 GB data. We vary the number of cores by booting the system with the kernel option $maxcpus=n$ where $n=2, 4, 6 $ and 8. As shown, when the degree of pipelines increases, the CPU usage grows as well,  due to the increased number of threads.

\begin{figure*}[htp]
\begin{minipage}[b]{0.5\linewidth}
\centering
\includegraphics[scale=0.5]{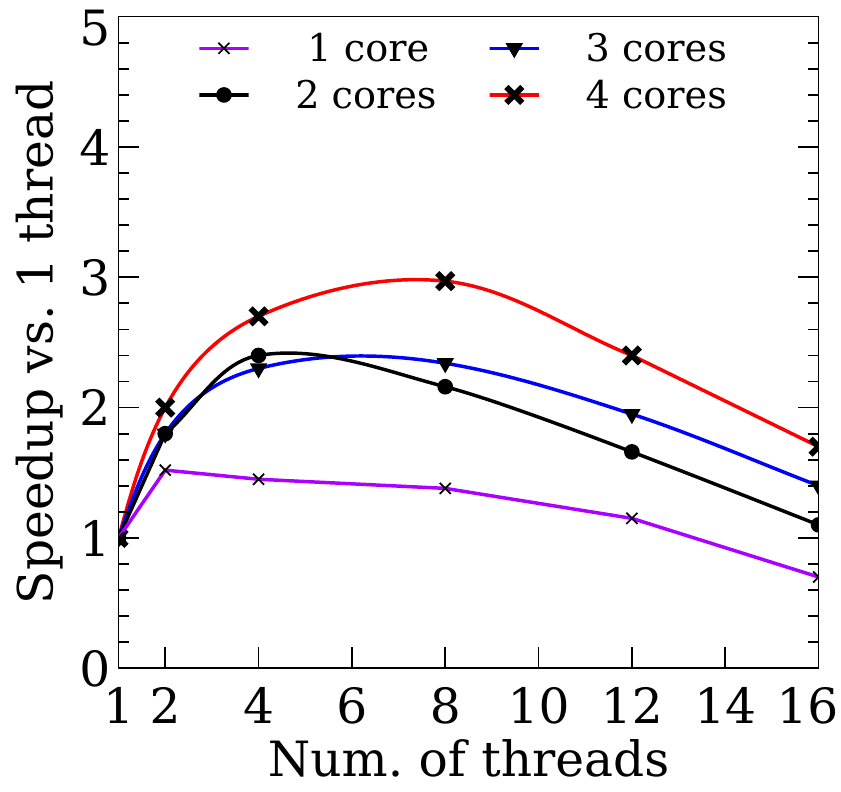}
\caption{Speedup of using multi-threading}
\label{fig:scalepipelinedegree}
\end{minipage}
\begin{minipage}[b]{0.5\linewidth}
\centering
\includegraphics[scale=0.5]{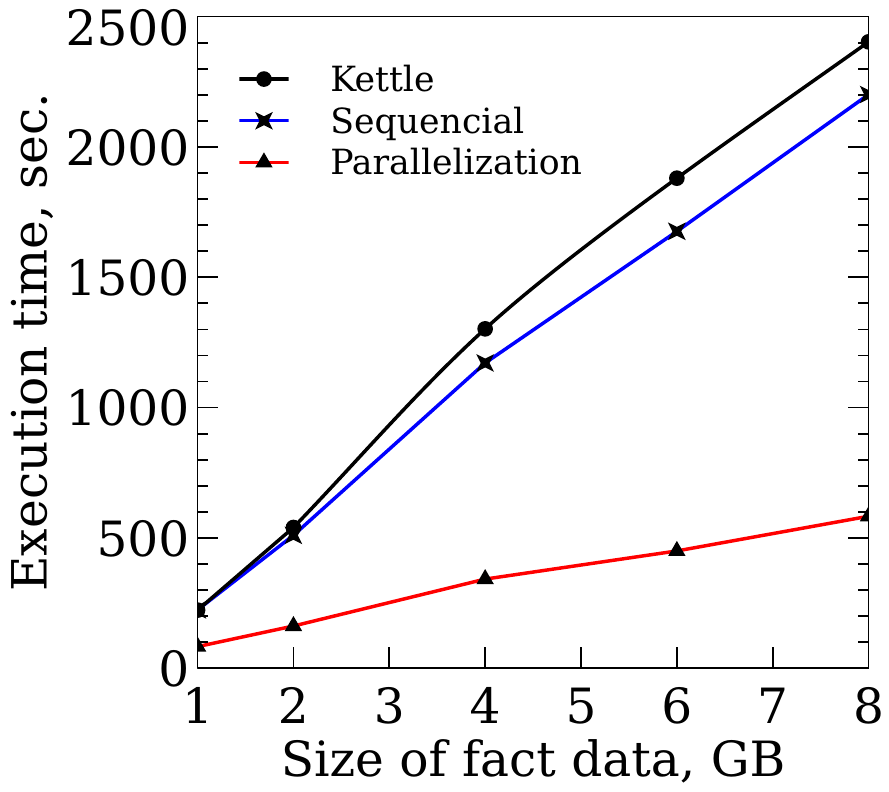}
\caption{Compared with Kettle}
\label{fig:total_optimized}
\end{minipage}
\end{figure*}

We now evaluate the inside-component parallelization using multi-threading. For testing purpose, we remove the index on the attribute \texttt{s\_nation} of the \texttt{supplier} table, and also disable the pipeline parallelization.  This leads to the lookup operation on the \texttt{supplier} table  be the bottleneck of the whole dataflow. We execute the dataflow when the fact table is loaded with 4 GB  data, and  the system is configured with 2, 4, 6 and 8 cores. Fig~\ref{fig:scalepipelinedegree} shows the speedup lines when  the number of the threads used by the lookup component  of \texttt{supplier} is scaled from 1 to 16.  The speedup of each test grows at the beginning but decreases when the number of threads is increased. The performance deterioration is also due to the computing resource competition for over threading.  The results also indicate that the speedup is better if  multi-core is used.

We now measure the performance improvement after applying the proposed optimization techniques.  We first test the sequential execution with and without using shared caching scheme, then test the pipeline parallelization. The number of pipelines is set to eight, which showed the best performance in the previous experiment.  The size of fact data is scaled from 1 to 8 GB, and  Figure~\ref{fig:total_optimized} shows the results. As shown, the sequential execution of using shared caching scheme has made $\sim10\%$ performance improvement, due to the removal of data copying operation in the execution trees. Thus, it was more efficient. When the pipeline parallelization is enabled, the execution  is 4.7x and 3.9x faster than the sequential executions without and with using the shared caching scheme, respectively. 

\subsection{Performance comparison with other ETL tools}
\begin{figure*}[htp]
\begin{minipage}[b]{0.5\linewidth}
\centering
\epsfig{file=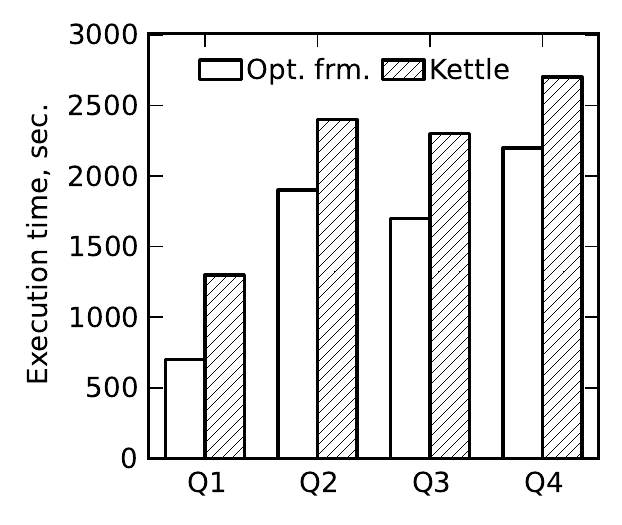, scale=0.7}
\caption{Sequential execution}
\label{fig:compared_others_seq}
\end{minipage}
\begin{minipage}[b]{0.5\linewidth}
\centering
\epsfig{file=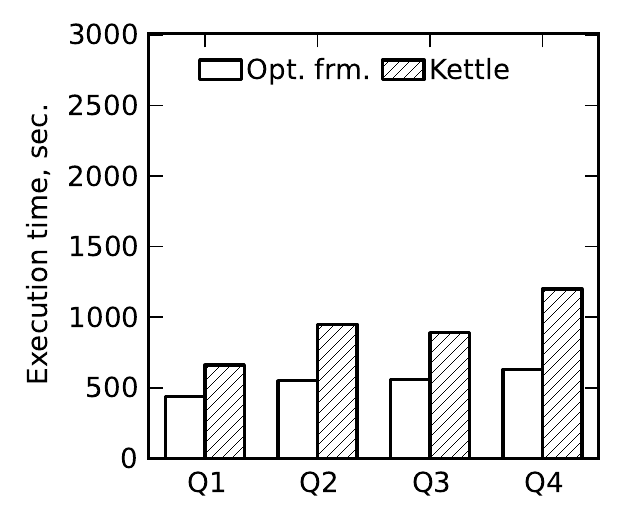, scale=0.7}
\caption{Horizontal parallelization}
\label{fig:compared_others_parallel}
\end{minipage}
\end{figure*}

We now compare the optimization framework (abbreviated as {\em Opt. frm.}) with the open source ETL tool, Kettle \cite{ketl2013}. We select the first query in each of the four SSB query categories  (note that the queries in each category only differentiate in the query conditions), denoted by Q1, Q2, Q3 and Q4. We run the test  with the fact table loaded 8 GB data set. Kettle also supports multi-threading parallelization within a component. To make the comparison fair, we use the same number of threads and the same step for both tools (eight threads for each query).  We first compare the sequential execution when the multi-threading parallelization is enabled.  The results  in Figure~\ref{fig:compared_others_seq} have shown the proposed optimization framework outperforms Kettle for all the four queries, due to the use of shared caching scheme. We now compare the pipeline parallelization to Kettle. Since Kettle does not provide the built-in supports for pipeline parallelization as ours, we have to make the workaround, \ie, for the connected row-synchronized components we horizontally split the dataflow into multiple parallel sub-flows by a splittor, then merge the results of all the sub-flows. The number of sub-flows is set to the same size of the pipelines of our framework (also set to eight). The results in Figure~\ref{fig:compared_others_parallel}   provide the further evidence that the pipeline parallelization for partitioned dataflows  is effective.

\section{Related Work}
\label{sec:relatedwork}
The optimization of ETL process is of particular importance, since ETL processes have to complete their tasks within specific time frames \cite{Vassiliadis2011}. Some research works are found to optimize the ETL process. Simitsis \et propose the theoretical framework \cite{simitsis2010,simitsis2005,simitsis20051} that formalizes an ETL scenario as a directed acyclic graph (DAG), and  handles ETL optimization as a state-space problem. When given an ETL scenario, the proposed framework can find an equivalent logical scenario that has the best execution time by searching the state space. This formalized DAG provides the theoretical foundation to the dataflow partitioning of our work. Tziovara \et propose the approach \cite{Tziovara20051} of producing the physical scenario with a logical ETL template as the input. But, unfortunately the the above research works do not have any open source implementations available for our comparison. Li and Zhan determine the critical path of a workflow with regard to the execution time. They analyzes the dependencies of tasks in a workflow and optimize the tasks without dependencies using parallelization \cite{LZ05}. Behrend and J\"{o}rg use rule-based optimization for the incremental ETL flows \cite{BJ10}. The rule-based optimization rewrites the rules with regards to the algebraic equivalences, however the rules need to be hard-coded during initial deployment \cite{Pedro2009}. The QoX-driven  approach \cite{Simitsisqos10,Dayal2009} is proposed to deal with the quality and the optimization objectives of ETL design. In comparison to the works above, we use the partitioning techniques based on the component types (in addition to parallelization) in order to optimize ETL dataflow. To the best of our knowledge, this is the first work to optimize ETL technology based on the partitioning techniques with respect to the component types existing in ETL dataflow.  

In the market a plethora of commercial ETL tools exist, such as IBM's Data Warehouse Manager \cite{IBMDWM}, Informatica's PowerCenter \cite{powercenter}, Microsoft's Data Transformation Services \cite{MDS} and Oracle's Warehouse Builder \cite{oracleDWbuilder}. The commercial tools provide the advanced GUI to ease ETL design, however most of them only provide the black-box optimization at logical level, \eg, re-ordering of the activities in the ETL dataflow. Some ETL engines such as PowerCenter \cite{powercenter} support ``push down" optimization, which pushes the operators that can be expressed in SQL from the ETL flow down to the source or target DBMS engine \cite{simitsis2005}. The remaining transformations are executed in the ETL data integration server. In contrast, our framework optimizes an ETL dataflow at three levels of granularity and uses pipelining and multi-threading based parallelization. This technique takes full advantage of computing resources, if the dataflow runs on a powerful machine.

One of the latest trends of ETL technologies is handling big data. One answer to this is Pig \cite{Olston2008}, an open source data flow system  for analyzing large data sets \cite{gates2009,olston2009}. Pig offers the SQL-style programming language, Pig Latin, which provides data manipulation constructs assembled in a dataflow.  Another similar system is Hive \cite{Thusoo2007},  built on top of Hadoop and also offering SQL-style language, HiveQL. In both systems, the programs  are compiled into sequences of MapReduce jobs, and executed in the Hadoop MapReduce environment.  Since they are designed to analyze big data for that reason they only have a limited ETL capabilities, somewhat like  DBMSs rather than full-fledged ETL tools. To complement this, Liu \et  propose a scalable dimensional ETL programming framework, ETLMR \cite{liu2011} using MapReduce, extended from  \cite{thomsen2009,thomsen2011}. ETLMR aims at the traditional RDBMS-based data warehousing system.  Furthermore, a dimensional ETL, CloudETL \cite{cloudetl2012}, is implemented for Hive  as the warehouse system. These two above mentioned ETL frameworks provide the built-in support for different data warehouse schemas including star, snowflake, and slowly changing dimensions. Only a few lines of code are needed to create a parallel ETL program \cite{liuvldb2012}. As opposed to the cloud-based ETL, the proposed framework aims at improving the efficiency of traditional ETL processes that typical run on a single server, whilst the cloud-based ETL improve the scalability, scaling out to many commodity servers and aims at dealing with the big data for today's data warehousing. The cloud-based ETL could be extended by using the proposed optimization techniques to improve the performance. We beleive that the use of the computing resource on each node could also be maximized by using the proposed optimization techniques, shared caching  and execution tree parallelization.

\section{Conclusion and Future Work}
\label{sec:conclusionandfuturework}

In order to improve the efficiency of an ETL dataflow, optimization is one of commonly used techniques, however, it is a non-trivial task due to the complexity of ETL technologies. In this paper, we have proposed a framework  to optimize ETL dataflow. The proposed framework first classifies the components based on their characteristics in the ETL dataflow, after that it partitions the dataflow at different granularities that are the foundations for the optimization.
Subsequently, we have proposed the concept of shared caching scheme. The use of shared cache not only minimizes the memory footprint of ETL dataflow but also reduces the CPU consumption to copy data, which is one of most frequently used operations in a dataflow. Furthermore, we have also presented the techniques to optimize the dataflow at different granularities by making use of  parallelization technologies, such as pipelining in execution trees and multi-threading at staggering components. In addition, we have proposed the formula and algorithm to estimate the optimal degree of pipelines in terms of the cost.  Finally, we have evaluated the proposed optimization framework comprehensively. The results show that the proposed framework is 4.7 times faster than the ordinary ETL dataflows (without using the proposed optimization techniques), and  outperforms the similar tool (Kettle).

In the future, the proposed optimization framework could be extended in several different directions. First, the self-adapt configuration is desirable such that system can automatically configure the number of pipelines needed, based on the statistics data of the previous runs. Second, it is interesting to add a cost model to the system, based on the cost model the system generates the optimal execution plan for ETL dataflow. Third, due to the growing interest of moving ETL to the cloud platform today, it is also interesting to improve the system so that it could be deployed in a clustering environment, whereas, the proposed optimization methods are used to take full advantage of the computing resources at each node.


\end{document}